\documentclass[twocolumn,preprintnumbers,amsmath,amssymb,prl,reprint,
groupedaddress,amsmath,citeautoscript,flushbottom,floatfix]{revtex4}
\usepackage{graphicx}
\usepackage{dcolumn}
\usepackage{bm}
\usepackage{amsmath}
\usepackage{amssymb}
\usepackage{amsmath}
\usepackage{booktabs}
\usepackage{multirow}
\usepackage{array}
\usepackage{color}
\makeatletter
\def\@hangfrom@section#1#2#3{\@hangfrom{#1#2#3}}
\makeatother
\newcommand{\vc}[1]{\boldsymbol{#1}}

\begin{document}

\title{Pseudo-Jahn-Teller Effect and Magnetoelastic Coupling
in Spin-Orbit Mott Insulators}
\author{Huimei Liu and Giniyat Khaliullin}
\affiliation{Max Planck Institute for Solid State Research,
Heisenbergstrasse 1, D-70569 Stuttgart, Germany}

\begin{abstract}
The consequences of the Jahn-Teller (JT) orbital-lattice coupling for
magnetism of pseudospin $J_{\rm eff}=1/2$ and $J_{\rm eff}=0$ compounds are
addressed. In the former case, represented by Sr$_2$IrO$_4$, this coupling
generates, through the so-called pseudo-JT effect, orthorhombic
deformations of a crystal concomitant with magnetic ordering. The
orthorhombicity axis is tied to the magnetization and rotates with it under
magnetic field. The theory resolves a number of puzzles in Sr$_2$IrO$_4$
such as the origin of in-plane magnetic anisotropy and magnon gaps,
metamagnetic transition, etc. In $J_{\rm eff}=0$ systems, the pseudo-JT effect
leads to spin-nematic transition well above magnetic ordering, which may
explain the origin of ``orbital order'' in Ca$_2$RuO$_4$.
\end{abstract}

\date{\today}
\maketitle

Electron-phonon coupling leads to a wide range of phenomena, from Cooper
pairing in metals to the Jahn-Teller (JT) effect in Mott insulators.
The JT effect, arising from coupling of the orbital degrees of freedom
of localized electrons to lattice vibrations (``orbital-lattice coupling''),
is a major source driving structural phase transitions.
Below the JT structural transition temperature $T_{\rm JT}$, the orbital
fluctuations are quenched, and resulting orbital order dictates
the spin-exchange couplings $J$ and magnetic structure below $T_{m}$
via so-called Goodenough-Kanamori rules~\cite{Goo63,Kug82}.
Typically, the JT and magnetic transitions are well separated; a canonical
example is LaMnO$_3$ with $T_{\rm JT}\sim 800$~K and $T_{m}\sim 140$~K.

The picture of successive orbital and spin orderings, and associated
Goodenough-Kanamori rules that guided spin-orbital physics in transition
metal compounds over decades, are based on a spin-orbital separation idea
assuming distinct energy scales and excitations in spin and orbital sectors.
Recently, materials based on late transition metal ions with strong spin-orbit
coupling (SOC) came into focus. In these compounds, spin-orbital separation
is no longer at work, and both magnetism and JT physics have to be
reformulated in terms of ``pseudospins''~\cite{Kha05}, or ``effective spins''
$J_{\rm eff}$~\cite{Kim08}, corresponding (but not always) to the total
angular momentum. While the pseudospin magnetism, especially
in $J_{\rm eff}=1/2$ systems, is now well understood (see the recent
reviews~\cite{Rau16,Win17,Tre17,Her18,Ber19}), the JT physics in spin-orbit
Mott insulators remains largely unexplored. Partially, this is due to the common
belief that JT coupling in $J_{\rm eff}=1/2$ systems is not essential at all,
since it cannot split the Kramers doublet.

In this Letter, we show that JT coupling has in fact a decisive impact on
low-energy magnetic properties of $J_{\rm eff}=1/2$, and even nominally
nonmagnetic $J_{\rm eff}=0$, compounds. By virtue of the pseudo-JT
effect~\cite{notePJT,Ber13,Ber06}, orbital-lattice coupling modulates
the spatial shape of the pseudospin
wave function and generates new terms in the Hamiltonian, describing the
pseudospin-lattice coupling. Albeit weak, these terms lead to the qualitative
effects: in the $J_{\rm eff}=1/2$ system Sr$_2$IrO$_4$, we predict that they
induce the tetragonal-to-orthorhombic structural transition, which turns
out to be instrumental for understanding the magnetic properties of this
compound, including metamagnetic behavior, the origin of magnon gaps, etc.
In $J_{\rm eff}=0$ systems, the JT coupling results in a simultaneous lattice
and spin-rotational symmetry breaking transition well above $T_m$.

\emph{Pseudospin-lattice coupling, $J_{\rm eff}=1/2$.---}
While physical ideas are generic to a broad class of spin-orbit Mott
insulators~\cite{Rau16,Win17,Tre17,Her18,Ber19,Liu18,San18}, we focus here on
Sr$_2$IrO$_4$, which is of special interest due to its quasi-two-dimensional
(2D) antiferromagnetism (AF)~\cite{Kim09} and magnon excitations~\cite{Kim12}
similar to those of La$_2$CuO$_4$~\cite{noteCu}.

The JT interaction operates in a quadrupolar channel; i.e., it couples
lattice deformations $\varepsilon_{\gamma}$ of certain symmetry $\gamma$ to
the orbital quadrupolar moments $Q^{\gamma}$ of valence electrons:
$\mathcal{H}_{\rm{JT}} \propto g_{\gamma}\varepsilon_{\gamma}Q^{\gamma}$.
Through the spin-orbit entanglement, this coupling should generate
pseudospin-lattice coupling $\mathcal{H}_{s-l}$ of the same form, with
$Q^{\gamma}$ replaced by the pseudospin quadrupoles $Q_{s}^{\gamma}$.
As no single-ion quadrupole can be formed out of pseudospin $S=1/2$,
$Q_{s}^{\gamma}$ should involve at least two sites, i.e., bilinear forms
${S_i^{\alpha}S_j^{\beta}}$ of a proper symmetry, suggesting a minimal coupling
$\mathcal{H}_{\rm{s-l}}^{ij}\propto \tilde{g}_{\gamma}\varepsilon_{\gamma}
Q_{s}^{\gamma}(ij)$. Below, we derive this interaction and evaluate the
coupling constants $\tilde{g}_{\gamma}$.

\begin{figure}
\begin{center}
\includegraphics[width=8.5cm]{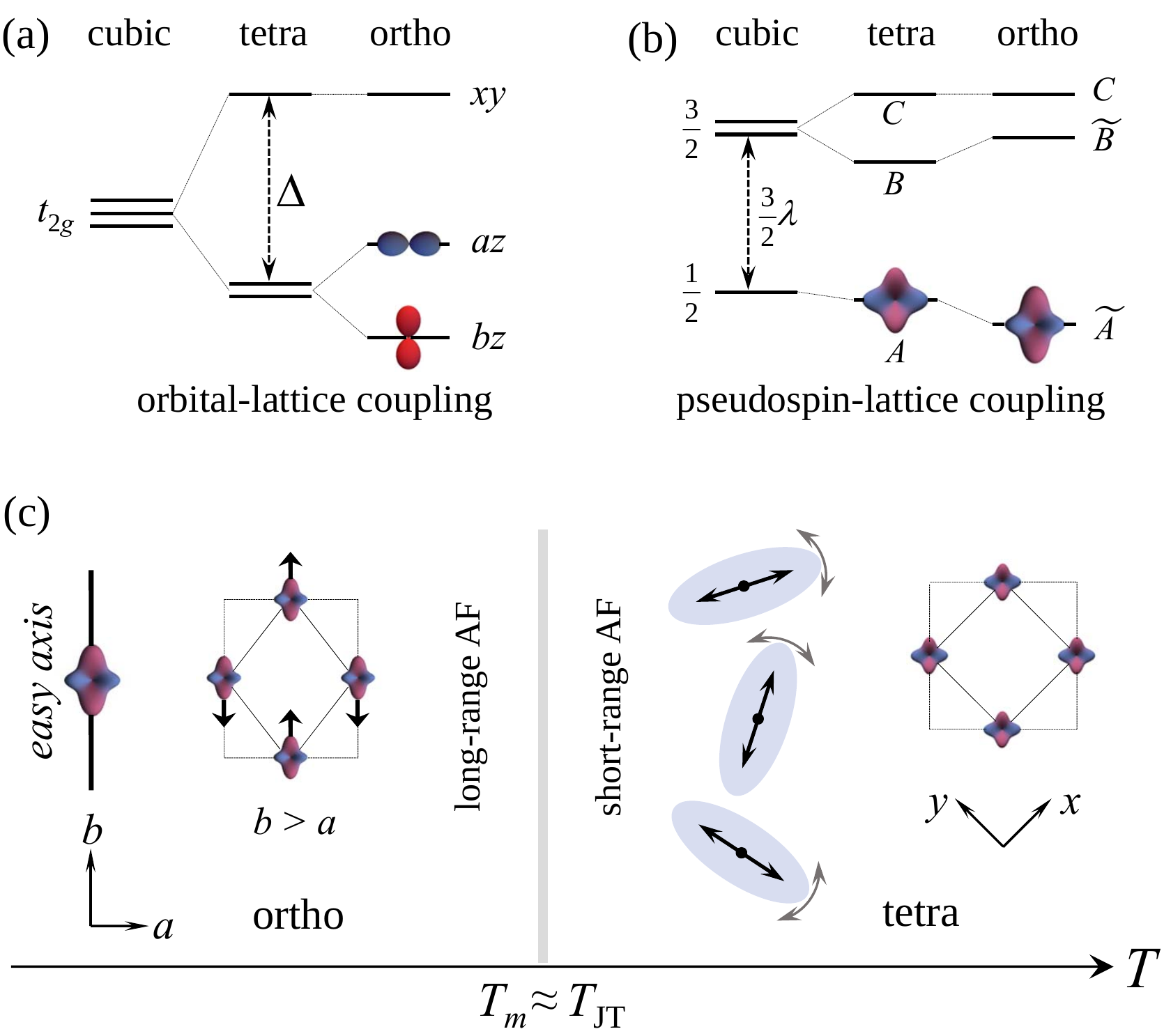}
\caption{
$t_{2g}$-hole level structure (a) without and (b) with SOC under cubic,
tetragonal, and orthorhombic crystal fields. ($\Delta>0$ corresponds to the
case of Sr$_2$IrO$_4$~\cite{Kim14,Fuj14,Bog15}). Elongation of a
crystal along the $b$ axis ($\varepsilon_1$ deformation) splits
$az$ (blue) and $bz$ (red) orbitals. This enhances the $bz$ component
of the ground state wave function $\tilde A$, breaking its tetragonal
symmetry (top view; $xy$ orbital is not shown for clarity).
(c) Illustration of the magnetoelastic coupling in Sr$_2$IrO$_4$.
Above the structural transition at $T_{\rm JT}\approx T_m$, symmetry is
tetragonal on average, but slowly rotating domains of the orthorhombic
distortions and quasi-2D magnetism develop. Below $T_{\rm JT}$, the tetragonal
symmetry is broken, selecting the $b$ axis for the moment direction.}
\label{fig:1}
\end{center}
\end{figure}

We consider the orthorhombic deformations which are common in
perovskites. In a tetragonal Sr$_2$IrO$_4$, these are $xy$ and $x^2\!-\!y^2$
type distortions, which we quantify by $\varepsilon_1\!=\!\frac{b-a}{b+a}$
and $\varepsilon_2\!=\!\frac{x-y}{x+y}$, correspondingly, using the
coordinate frames of Fig.~\ref{fig:1}, where $a$ and $b$ axes are rotated
by 45$^{\circ}$ with respect to cubic $x$ and $y$ axes.
$\varepsilon_1$ and $\varepsilon_2$
measure elongation of a crystal along $b$ and $x$ directions, respectively.
The distortions split $t_{2g}$ level via the JT coupling:
\begin{align}
\label{eq:JT}
\mathcal{H}_{\rm{JT}} = g_1\varepsilon_1\;(n_{az}\!-\!n_{bz}) +
g_2\varepsilon_2\;(n_{xz}\!-\!n_{yz}),
\end{align}
where $n_{az}=d_{az}^\dagger d_{az}$ and $n_{bz}=d_{bz}^\dagger d_{bz}$ are
densities of the $az=\frac{1}{\sqrt{2}}(x-y)z$ and
$bz=\frac{1}{\sqrt{2}}(x+y)z$ orbitals. This coupling mixes the Kramers
doublets $A$ and $B$ of Ir$^{4+}$ ion (Fig.~\ref{fig:1}), resulting in
the ``orthorhombically distorted'' pseudospin wave function $\tilde A$:
\begin{align}
\label{eq:gs}
|\widetilde{A}_{\pm}\rangle = \frac{1}{\sqrt{1\!+\!|\eta_{\pm}|^2}}
\left(\;|A_{\pm}\rangle + \eta_{\pm}|B_{\mp}\rangle\;\right),
\end{align}
where
$\eta_{\pm}=\frac{\cos\theta}{E_{BA}}(\pm ig_1\varepsilon_1+g_2\varepsilon_2)$.
The angle $\theta$ with $\tan 2\theta=2\sqrt{2}\lambda/(\lambda+2\Delta)$
quantifies a tetragonal field $\Delta$ relative to SOC constant $\lambda$,
and $E_{BA}\sim \tfrac{3}{2}\lambda$ is the energy difference between
$A$ and $B$ levels~\cite{noteAB}. The ``tetragonal'', i.e., unperturbed
wave functions $|A_{\pm}\rangle= \sin\theta\;|0,\pm\tfrac{1}{2}\rangle-
\cos\theta\;|\pm 1,\mp\tfrac{1}{2}\rangle$ and
$|B_{\pm}\rangle=|\pm 1,\pm\tfrac{1}{2}\rangle$,
in terms of $t_{2g}$ orbital and spin quantum numbers $|l_z,s_z\rangle$.

Next, we inspect how the shape distortions of the ground state wave function
$\tilde A$ affect the pseudospin interactions. Deformations are assumed to be
quasistatic (adiabatic approximation). Projecting the Kugel-Khomskii-type
spin-orbital Hamiltonian, Eq.~(3.11) of Ref.~\cite{Kha05}, onto $\tilde A$
subspace, we find
$\mathcal{H}\!=\!\mathcal{H}_{\rm{s}}\!+\!\mathcal{H}_{\rm{s-l}}$.
$\mathcal{H}_{\rm{s}}$ comprises the nearest-neighbor Heisenberg $J$,
Ising $J_z$, Dzyaloshinskii-Moriya $D$, and pseudodipolar $K$ terms
\begin{align}
\label{eq:s}
J\vec S_i\!\cdot\!\vec S_j\!+ \!J_zS_i^z S_j^z\!+\!
\vec D\! \cdot \![\vec S_i\!\times\!\vec S_j]\!+
\!K(\vec S_i\!\cdot\!\vec r_{ij})(\vec S_j\!\cdot\!\vec r_{ij})
\end{align}
derived earlier~\cite{Jac09}, while
\begin{equation}
\label{eq:s-l}
\mathcal{H}_{s-l}^{ij}\!=\!
\tilde{g}_1 \varepsilon_1\;(S_i^xS_j^y\!+\!S_i^yS_j^x) +
\tilde{g}_2 \varepsilon_2\;(S_i^xS_j^x\!-\!S_i^yS_j^y)
\end{equation}
constitutes the (pseudo)spin-lattice interaction that we are looking
for~\cite{SM}. It linearly couples the spin quadrupoles $Q_s^1$ and $Q_s^2$
of $xy$ and $x^2\!-\!y^2$ symmetries to corresponding lattice deformations.
In essence, $\mathcal{H}_{s-l}$ is nothing but
$\mathcal{H}_{\rm{JT}}$ ``reincarnated'' as a spin-lattice coupling in
$J_{\rm eff}=1/2$ insulator. The coupling constants $\tilde{g}$ are
renormalized from $g$ of Eq.~\ref{eq:JT} to $\tilde{g}\!=\!\kappa g$ by
$\kappa\!\simeq\!\frac{t^2}{U}\frac{\sin^2 2\theta}{E_{BA}}\frac{J_H}{U}$,
where $t$, $U$, and $J_H$ are hopping amplitude, Coulomb repulsion, and
Hund's coupling, respectively. Roughly, we estimate
$\kappa\sim 5\times 10^{-3}$ and hence $\tilde{g} \sim 25$~meV in
Sr$_2$IrO$_4$, using $g\sim 5$~eV typical for $t_{2g}$ systems.
In $J_{\rm eff}=1/2$ compounds based on $4d$ Ru$^{3+}$ and $3d$ Co$^{2+}$
ions, $\kappa$ and $\tilde{g}$ should increase as $1/\lambda$.

\emph{Breaking tetragonal symmetry.---}
Having derived spin-lattice interaction $\mathcal{H}_{s-l}$, we
discuss now its consequences for low-energy properties of Sr$_2$IrO$_4$.
First of all, just as the JT coupling, it should lead to the structural
instability as soon as the spin quadrupolar moments $Q_{s}^{\gamma}$ develop
within the (quasi) long-range ordered magnetic domains. Denoting the
staggered moment direction by $\alpha$, $\vec n=S (\cos\alpha, \;\sin\alpha)$,
we find $\langle Q_{s}^1\rangle=-S^2\sin2\alpha$ and
$\langle Q_{s}^2\rangle=-S^2\cos2\alpha$ per bond. From Eq.~(\ref{eq:s-l}) and
elastic energy $\tfrac{1}{2}K_\gamma\varepsilon_\gamma^2$, the spin-lattice
induced orthorhombic deformations follow:
\begin{align}
\label{eq:def}
\langle\varepsilon_1\rangle\!=\!\frac{\Gamma_1}{\tilde{g}_1}\sin2\alpha, \;\;
\langle\varepsilon_2\rangle\!=\!\frac{\Gamma_2}{\tilde{g}_2}\cos2\alpha,
\end{align}
where $\Gamma_\gamma=2 S^2 \tilde{g}_\gamma^2/K_\gamma$.
A mean-field part of $\mathcal{H}_{s-l}$ (4) reads then as follows:
\begin{equation}
\label{eq:G}
\Gamma_1\sin2\alpha\;(S_i^xS_j^y\!+\!S_i^yS_j^x) +
\Gamma_2\cos2\alpha\;(S_i^xS_j^x\!-\!S_i^yS_j^y),
\end{equation}
with $\alpha$ to be obtained by minimizing the ground state energy $E_\alpha$.
Classically,
$E_\alpha \!=\!\rm{const} + S^2(\Gamma_1\!-\!\Gamma_2)\cos^2 2\alpha$~\cite{noteK}.
For $\Gamma_1>\Gamma_2$, $E_\alpha$ is minimized at $\alpha=45^{\circ}$,
which is exactly the case of Sr$_2$IrO$_4$~\cite{Cao98,Kim09}. Our theory
predicts then $\varepsilon_1$-type ($b>a$) orthorhombic distortion, as
depicted in Fig.~\ref{fig:1}(c). This type of distortion is natural for
perovskites, as it does not affect the Me-O-Me bond length.

Breaking $C_4$ symmetry by spin-lattice coupling opens the in-plane magnon
gap already on a level of linear spin-wave theory. Equations.~\ref{eq:s}
and~\ref{eq:G} give $\omega_{ab}\!\simeq\!8S\sqrt{J\Gamma_1}$. With
$\omega_{ab}\!\sim\! 2.1-2.4$~meV~\cite{Gim16,Gre17} and
$J\!\sim\!100$~meV~\cite{Kim12,Fuj12}, we evaluate $\Gamma_1\!\sim\!3\;\mu$eV.
Equation.~\ref{eq:def} predicts then the spin-lattice induced distortion
of the order of $\varepsilon_1\sim 10^{-4}$~\cite{noteCo}. The twofold
$C_2$ anisotropy of magnetoresistivity~\cite{Wan15} and the signatures of
orthorhombic distortions~\cite{Dhi13,Ye13} in Sr$_2$IrO$_4$ find a natural
explanation within our theory. Future experiments using, e.g., Larmor
diffraction~\cite{Naf16} should be able to quantify $\varepsilon_1$ directly.
We note also that the deformation induced magnon gap $\omega_{ab}$ far exceeds
interlayer couplings~\cite{Tak16}, and should therefore be essential
for establishing the magnetic order at high $T_m\sim 240$~K.

To summarize up to now, the combined action of spin-orbit and JT couplings
results in the interaction between magnetic quadrupoles and lattice
deformation. Dynamically, coupled oscillations of the $\vec n$-moment
direction and lattice vibrations (magnetoacoustic effects~\cite{Kit58,Tur83})
are expected; this is an interesting topic for future research. Most
importantly, a structural instability is inevitable no matter how large
SOC is; this invalidates a common assertion that high tetragonal symmetry
of $J_{\rm eff}=1/2$ system Sr$_2$IrO$_4$ is protected by large SOC.

\begin{figure}
\begin{center}
\includegraphics[width=8.5cm]{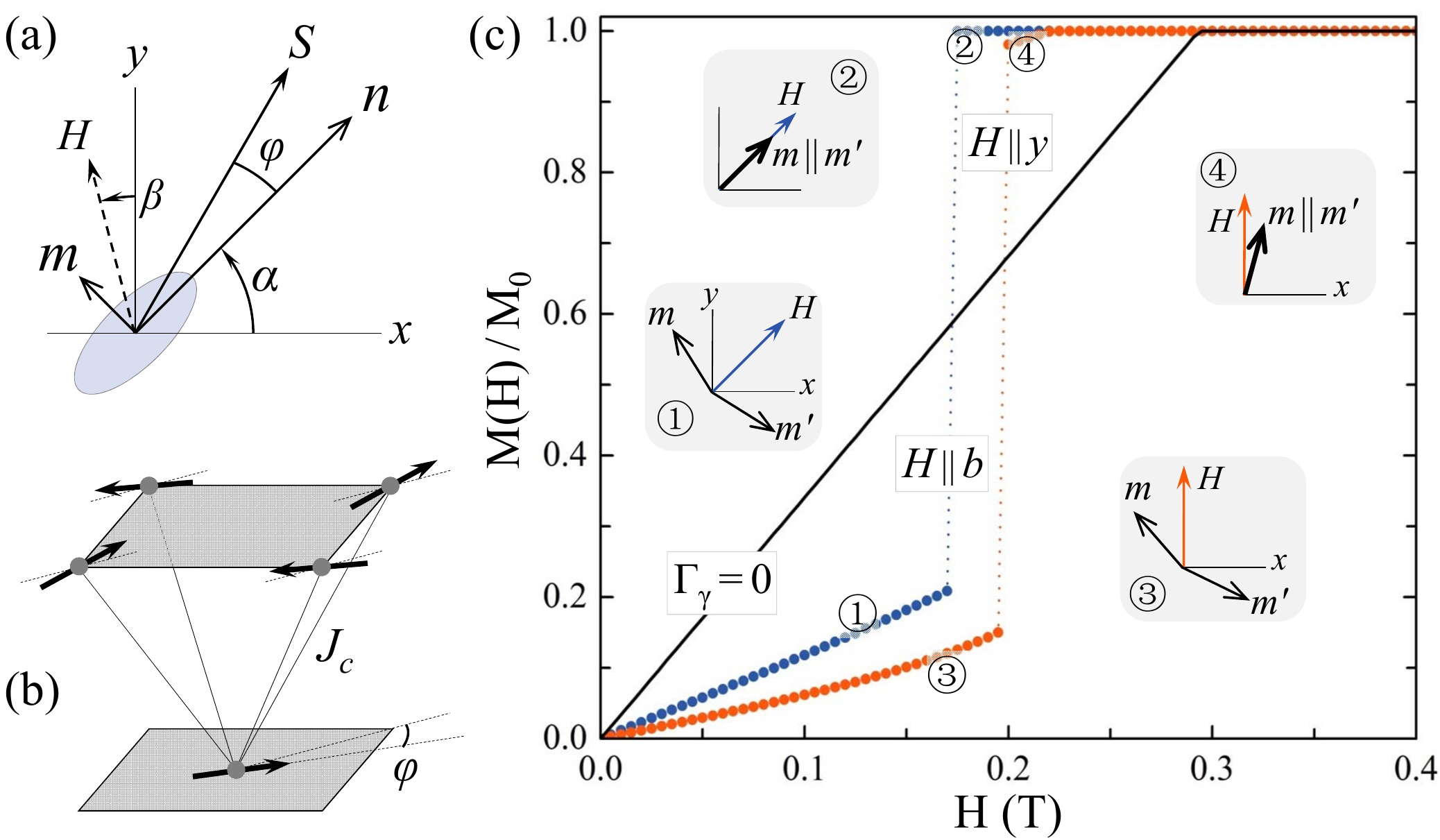}
\caption{Schematic of (a) staggered $\vec n$ and canted $\vec m$ moments,
magnetic field $H$, and spin-lattice induced orthorhombic deformation
(shaded ellipse), and (b) interlayer AF coupling $J_c$ between spins.
(c) Magnetization curves for a magnetic field applied along the ''easy''
($H \parallel b$) and ''hard'' ($H \parallel y$) axes. While the
magnetization $\vec{M}\propto\vec{m}+\vec{m}^\prime$ grows linearly with $H$
when $\Gamma_\gamma=0$, a metamagnetic transition caused by magnetoelastic
coupling is observed at finite $\Gamma_\gamma$ (we used $\Gamma_1=3\;\mu$eV and
$\Gamma_2=0.6\Gamma_1$). Insets depict the mutual orientation of $\vec{m}$
and $\vec{m}^\prime$ moments on different layers at representative points
on the $M(H)$ curves.}
\label{fig:2}
\end{center}
\end{figure}

\emph{Metamagnetic transition, in-plane magnon gap}.---
We discuss now further manifestations of magnetoelastic coupling in
Sr$_2$IrO$_4$. Via spin-lattice coupling, the reorientations of moments under
external magnetic field will affect lattice deformations. The latter, in turn,
modifies the magnetic anisotropy potential. Such feedback effects result in
a nonmonotonic behavior of magnetization $M(H)$. In Sr$_2$IrO$_4$, spins are
canted by angle $\varphi\simeq D/2J \sim 12^{\circ}$~\cite{Bos13}, see
Figs.~\ref{fig:2}(a) and \ref{fig:2}(b). Magnetic field couples to the canted moments $\vec{m}$.
To calculate $M(H)$, we use a simple model in Fig.~\ref{fig:2}(b) for the
interlayer coupling. The total energy $E$ depends now on two angles $\alpha$
and $\alpha^\prime$, corresponding to the moment directions in different
layers, and the field direction $\beta$. We find
$E(\alpha,\alpha^\prime,\beta)=\rm{const}+\tfrac{S}{2}F$, with
\begin{eqnarray}
\label{eq:F}
F &\!= \sin\varphi [h_c\cos(\alpha\!-\!\alpha^\prime)\!-
\!h\cos(\alpha\!-\!\beta)\!-\!h\cos(\alpha^\prime\!-\!\beta)]\!
\notag \\
&-\tfrac{S}{2}[\Gamma_1(\sin 2\alpha\!+\!\sin 2\alpha^\prime)^2
\!+\!\Gamma_2(\cos 2\alpha\!+\!\cos 2\alpha^\prime)^2].
\end{eqnarray}
Here, $h=g\mu_{B}H$, and $h_c=4J_cS\sin\varphi$ is the interlayer field.
Minimization of $F$ gives $\alpha$ and $\alpha^\prime$ as a function of
$\vec H$, from which the canted moments $\vec m$ and $\vec m^\prime$ on
different planes and total magnetization $\vec M$ follow. The deformations
$\varepsilon_1$ and $\varepsilon_2$ are given by Eq.~(\ref{eq:def}), where
$\sin 2\alpha$ and $\cos 2\alpha$ replaced now by
$\tfrac{1}{2}(\sin 2\alpha\!+\!\sin 2\alpha^\prime)$ and
$\tfrac{1}{2}(\cos 2\alpha\!+\!\cos 2\alpha^\prime)$, respectively;
this implies the field dependence of the deformations (magnetostriction).

Fig.~\ref{fig:2}(c) shows $M(H)/M_0$ calculated with $h_c =18\;\mu$eV
($\simeq 0.16$~T). Without spin-lattice coupling, $\vec m$ and $\vec m^\prime$
gradually rotate towards each other and $M$ grows monotonically.
Spin-lattice induced anisotropy results in a metamagnetic transition as
observed~\cite{Cao98,Kim09}. At $H=H_{cr}$, $\vec m$ and $\vec m^\prime$ flip
and become parallel. For $\Gamma_1>\Gamma_2$ as in Sr$_2$IrO$_4$, $H_{cr}$ for
easy-axis $b$ is lower than that for hard axis; this result has recently been
confirmed experimentally~\cite{Por18}. We note that $M(H)$ near $H_{cr}$
is sensitive to angle $\beta$, so the quenched disorder and sample
alignment issues should be relevant in the data analysis.

\begin{figure}
\begin{center}
\includegraphics[width=8.5cm]{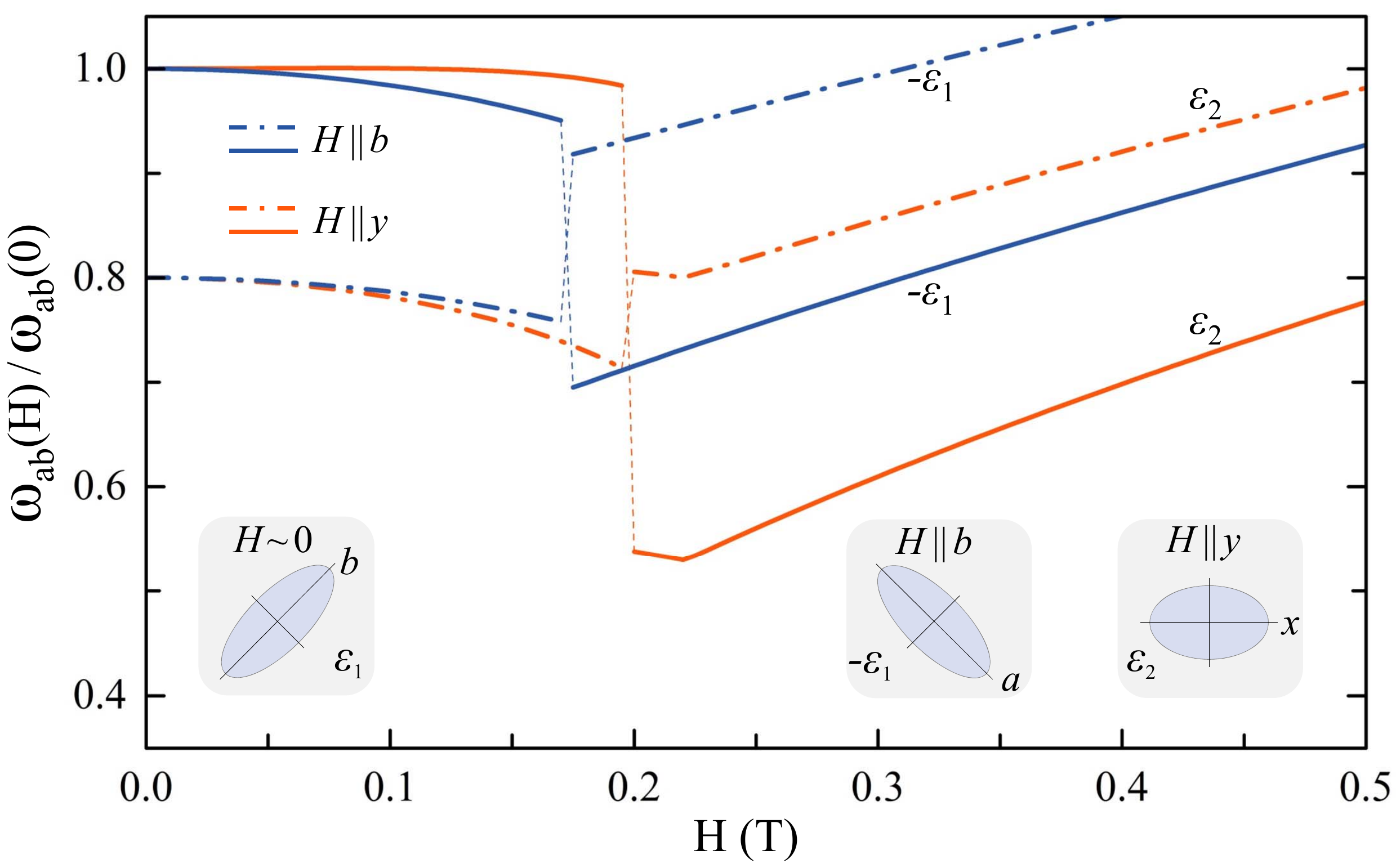}
\caption{Magnon gaps as a function of magnetic field along $b$ ($[010]$ in
orthorhombic notation, blue) and $y$ ($[110]$, red) directions.
Dash-dotted (solid) line corresponds to the in-phase (antiphase) rotations
of ${\vec m}$ and ${\vec m}^\prime$ moments. At small $H$, the distortion is
of $\varepsilon_1$ symmetry. At large $H$, it remains $\varepsilon_1$ type for
$H\parallel b$. For $H\parallel y$, the deformation changes from large
$\varepsilon_1$ to small $\varepsilon_2$, resulting in a drop of the anisotropy
energy and $\omega_{ab}$. The parameters are as in Fig.~\ref{fig:2}.}
\label{fig:3}
\end{center}
\end{figure}
Next, we discuss the in-plane magnon gaps generated by spin-lattice
coupling $\mathcal{H}_{s-l}$. Because of interlayer coupling, there are two
different modes. At small fields, $H\ll H_{cr}$, the optical and acoustic mode
gaps are $8S\!\sqrt{J(\Gamma_1 \!+\! \tfrac{\sin\varphi}{4S} h_c)}$ and
$8S\!\sqrt{J\Gamma_1}$, respectively. Above the metamagnetic transition,
$H \geq H_{cr}$, we find
\begin{align}
\omega_{ab}^\pm \!\simeq\!8S\!\sqrt{J\{\Gamma(\alpha)\!+\!\tfrac{\sin\varphi}{8S}
[h\cos(\alpha\!-\!\beta)-h_c \pm h_c]\}} \;.
\label{eq:w}
\end{align}
Here, $\Gamma(\alpha)=\Gamma_1\sin^22\alpha + \Gamma_2\cos^22\alpha$, and
$\alpha$ follows from
$2S(\Gamma_1\!-\!\Gamma_2)\sin4\alpha=h\sin\varphi\sin(\alpha\!-\!\beta)$.
For $H\parallel b$, this gives $\alpha\!=\!\beta\;(=\!-\tfrac{\pi}{4})$ and
$\Gamma(\alpha)=\Gamma_1$. For $\vec H$ along $y$ axis,
$\alpha\!\sim\!\beta\;(=0)$ and thus $\Gamma(\alpha)\sim\Gamma_2$; this implies
weak distortion $\varepsilon_2$ and smaller magnon gap. The main message is
that the magnon gaps become strongly dependent on the field direction, as
shown in Fig.~\ref{fig:3}. The above equations should help to quantify
$\Gamma_1$ and $\Gamma_2$ from experiments. The results in
Fig.~\ref{fig:3} are qualitatively consistent with the recent Raman
data~\cite{Gim16}; a detailed analysis would require a derivation of the
Raman matrix elements necessary for the mode assignment.

Via the magnetoelastic coupling, quasi-2D AF correlations above
$T_m$~\cite{Fuj12,Val15} should lead to slowly fluctuating lattice
deformations (see Fig.~\ref{fig:1}) which, in turn, will affect phonon
dynamics. Indeed, strong Fano anomalies of phonons have been observed in
Sr$_2$IrO$_4$~\cite{Gre16}.

\emph{Spin-nematic order in $J_{\rm eff}=0$ systems.---}
Finally, we move to pseudospin $J_{\rm eff}=0$ case, and show that, despite
having neither orbital nor spin degeneracy, the JT coupling is relevant even
here. In general, the $J_{\rm eff}=0$ compounds are of interest because they host
``excitonic'' magnetism~\cite{Kha13} - magnetic order via condensation
of spin-orbit $J_{\rm eff}=0\!\rightarrow\!1$ excitations. The expected
non-Heisenberg-type magnon and amplitude (Higgs) modes have been observed in
Ca$_2$RuO$_4$~\cite{Jai17,Sou17}. Also, $J_{\rm eff}=0$ systems illustrate well
the interplay between three ``grand forces'' in Mott insulators - the JT
coupling, spin-orbital exchange interaction, and spin-orbit
coupling~\cite{Kha05}.

As a toy model, we consider 2D square lattice of $J_{\rm eff}=0$ ions
(e.g., $d^4$ Ru$^{4+}$) in an octahedral field. The $t_{2g}^4$ orbital
configuration is subject to the JT effect; however, it is opposed by SOC
that favors spin-orbit singlet $J_{\rm eff}=0$ instead~\cite{Kha13,Mee15}.
This competition can be resolved by mixing the $J_{\rm eff}=0$ wave function
with the excited $J_{\rm eff}=1$ states, by virtue of spin-orbital exchange
interactions. Since $J_{\rm eff}=1$ level hosts a quadrupolar moment, the
ground state becomes JT active, and the phase transition, breaking
simultaneously the lattice and spin-rotational symmetries, may develop.
In essence, this is the ``spin nematic'' phase discussed in the context of
large $J_{\rm eff}$ systems~\cite{Che10}, but with the quadrupolar order
parameter depending now on the $J_{\rm eff}=1$ fraction in the condensate.

A minimal model for $d^4$ system can be cast in terms of bosons
$\vc T=(T_x,T_y,T_z)$, describing excitations from the ground state
$J_{\rm eff}=0$ singlet to $J_{\rm eff}=1$ triplet. The spin-orbit $\lambda$ and
exchange $J\!\simeq\!4t^2/U$ couplings read, in a cubic limit,
as follows~\cite{Kha13}:
\begin{align}
\mathcal{H}_{\lambda, J}=\lambda\sum_in_i^T +J\sum_{<ij>}
\tfrac{1}{4}(\vc T^\dagger_i\!\cdot\!\vc T_j-\vc T_i\!\cdot\!\vc T_j+H.c.),
\end{align}
where $n^T=n_x^T+n_y^T+n^T_z$ and $n_x^T=T_x^{\dagger}T_x$, etc. The $T$ bosons
are subject to ``hard-core'' constraint $n_i^T \leq 1$ which we treat on
a mean-field level~\cite{Som01,Mat04,SM}.

\begin{figure}
\begin{center}
\includegraphics[width=8.5cm]{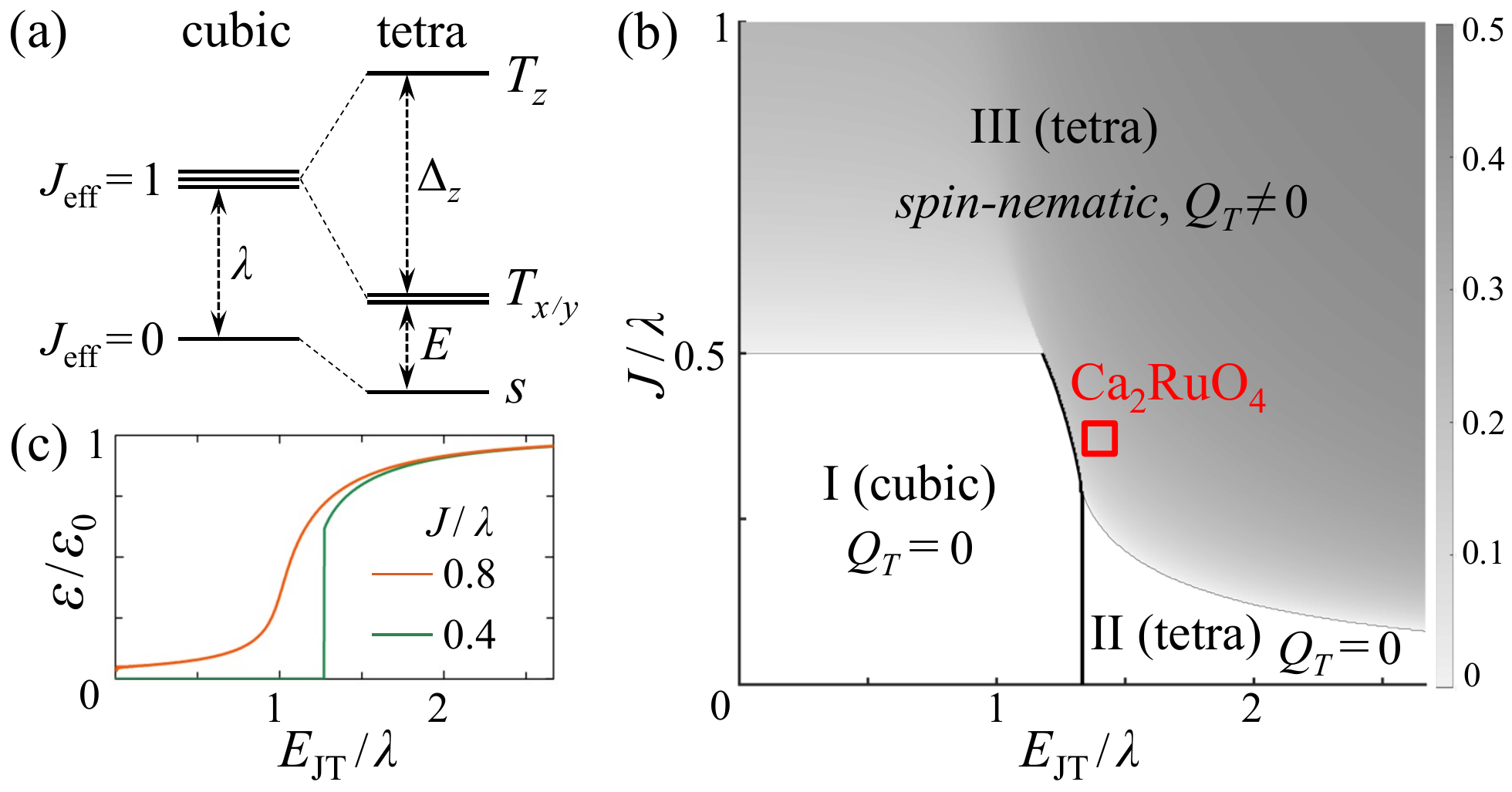}
\caption{(a) Singlet-triplet level structure under tetragonal distortion.
(b) Phase diagram of $J_{\rm eff}=0$ system. Small $J$ area contains two
nonmagnetic phases separated by a first order transition (thick line).
In phase I, the JT effect is fully suppressed, while phase II is tetragonally
distorted. As $J$ increases, the exchange interactions promote condensation of
the $T_{x/y}$ states, forming spin-nematic phase with nonzero $Q_T$ moment
(quantified by color intensity) and $XY$-type magnetism.
(c) Lattice distortion $\varepsilon$ relative to its value $\varepsilon_0$
at $\lambda=0$ for different $J$ couplings.}
\label{fig:4}
\end{center}
\end{figure}

We consider now a tetragonal distortion $\varepsilon\!=\!\frac{x+y-2z}{x+y+z}$
of RuO$_6$ octahedra. The JT coupling splits the $xy$ and $xz/yz$ orbital
levels by $g\varepsilon$:
$\mathcal{H}_{\rm{JT}}=g\varepsilon\;\tfrac{1}{3}(n_{zx}\!+\!n_{yz}-\!2n_{xy})$.
This coupling modifies a single-ion level structure of Ru$^{4+}$ as shown in
Fig.~\ref{fig:4}(a), such that $J_{\rm eff}=1$ triplet splits into $T_{x/y}$
doublet and $T_z$ singlet by
$\Delta_z(\delta)=(\delta+\sqrt{1+\delta^2}-1)\lambda$,
where $\delta=g\varepsilon/2\lambda$. As a result, the spin gap reduces
from $\lambda$ to $E(\delta)=
(\frac{1}{2}+\sqrt{\frac{9}{4}-\delta+\delta^2}-\sqrt{1+\delta^2})\lambda$.
At the critical value of $E(\delta)\sim J$, the $T_{x/y}$ doublet condenses,
forming a ground state with finite quadrupole moment $Q_T=n_x^T+n_y^T-2n^T_z$.
While the cubic symmetry may be broken at finite temperature $T_{\rm JT}$,
long-range magnetic order is delayed due to $XY$-type phase fluctuations;
therefore, $T_m$ and $T_{\rm JT}$ are separated in quasi-2D $J_{\rm eff}=0$
systems. We think that the ''orbital order'' in Ca$_2$RuO$_4$ near
260 K~\cite{Zeg05}, well above $T_m$, is in fact the JT driven
spin-nematic order. The observed $XY$-type magnons~\cite{Jai17} further
support the picture of spin-orbit entangled $T_{x/y}$ condensate.

A mean-field phase diagram of $\mathcal{H}_{\lambda,J}+\mathcal{H}_{\rm{JT}}$,
supplemented by the elastic energy $\tfrac{1}{2}K\varepsilon^2$, is shown in
Fig.~\ref{fig:4}(b) as a function of $J/\lambda$ and $E_{\rm JT}/\lambda$.
$E_{\rm JT}=\Delta/3$ is the JT stabilization energy, where
$\Delta=\frac{2g^2}{3K}$ is the $t_{2g}$ orbital splitting at $\lambda=0$. At
small $J$ and $E_{\rm JT}$, SOC imposes the $J_{\rm eff}=0$ phase I; at large
$E_{\rm JT}$, it gives way to the JT-distorted nonmagnetic phase II with finite
spin-gap $E$. In phase III, stabilized by a cooperative action of the exchange
and JT couplings, $XY$-type magnetic condensate is formed, which, in turn,
helps to recover the JT distortion [see Fig.~\ref{fig:4}(c)].

Interestingly, the observed magnon bandwidth $\sim 2J\sim 50$~meV~\cite{Jai17}
and ratio $\Delta/2\lambda\sim 2$~\cite{Das18,Gre18} locate Ca$_2$RuO$_4$ in
the critical area of the phase diagram [see Fig.~\ref{fig:4}(b)]. This suggests
that an unusual magnetism~\cite{Jai17,Sou17} and extreme sensitivity of
Ca$_2$RuO$_4$ to external perturbations~\cite{Sow17,Nak13} are caused by
frustration among the JT, spin-orbit, and exchange interactions, further
boosted by its proximity to metal-insulator transition.

To conclude, in contrast to the common wisdom, the JT coupling remains an
essential part of the low-energy physics in spin-orbit $J_{\rm eff}=1/2$, and
even $J_{\rm eff}=0$, Mott insulators. Converted into pseudospin-lattice
coupling via spin-orbit entanglement, it leads to the structural transitions
and magnetoelastic effects. We have shown that the JT coupling resolves
hitherto unexplained puzzles of $J_{\rm eff}=1/2$ Sr$_2$IrO$_4$, and
is essential for the phase behavior of $J_{\rm eff}=0$ Ca$_2$RuO$_4$. This
leads us to believe that pseudospin-lattice coupling should be generic to a
broad class of spin-orbit $J_{\rm eff}$ compounds, including the Kitaev-model
materials of high current interest~\cite{Rau16,Win17,Tre17,Her18}. In the
latter, the pseudospins are highly frustrated, and their coupling to lattice
may lead to more radical effects than in conventional, unfrustrated magnets
like Sr$_2$IrO$_4$.

\acknowledgments
We thank B. J. Kim, J. Porras, J. Bertinshaw, B. Keimer, J. Chaloupka,
and O. P. Sushkov for discussions. We acknowledge support by the
European Research Council under Advanced Grant No. 669550 (Com4Com).


\clearpage
\onecolumngrid
\begin{center}
\textbf{\large Supplemental Material for\\
Pseudo Jahn-Teller Effect and Magnetoelastic Coupling
in Spin-Orbit Mott Insulators}
\end{center}
\setcounter{equation}{0}
\setcounter{figure}{0}
\setcounter{table}{0}
\setcounter{page}{7}
\makeatletter
\renewcommand{\thesection}{S\Roman{section}}
\renewcommand{\thetable}{S\arabic{table}}
\renewcommand{\theequation}{S\arabic{equation}}
\renewcommand{\thefigure}{S\arabic{figure}}

\section{A. Derivation of pseudospin-lattice coupling
Hamiltonian $\mathcal{H}_{\rm s-l}$}

For $t_{2g}$ orbital system with spin $s=1/2$ (single-hole $d^5$ or
single-electron $d^1$ case), Kugel-Khomskii type spin-orbital exchange in
a perovskite lattice can be written as~\cite{Kha05}:
\begin{align}
\mathcal{H}_{ij}^{(\gamma)} &=
\frac{2t^2}{E_1} \left(\vec s_i\vec s_j+\tfrac{3}{4} \right)
\left(\mathcal{O}^{(\gamma)}_{ij}-\tfrac{1}{2}n_i^{(\gamma)}-
\tfrac{1}{2}n_j^{(\gamma)} \right) \notag \\
        &+\frac{2t^2}{E_2} \left(\vec s_i \vec s_j-\tfrac{1}{4} \right)
\left(\mathcal{O}^{(\gamma)}_{ij}+\tfrac{1}{2}n_i^{(\gamma)}+
\tfrac{1}{2}n_j^{(\gamma)} \right) \notag \\
        &+\left(\frac{2t^2}{E_3}-\frac{2t^2}{E_2} \right)
\left(\vec s_i \vec s_j-\tfrac{1}{4}\right)\tfrac{2}{3}\mathcal{P}^{(\gamma)}_{ij}.
\label{eq:H1}
\end{align}
Here, $t$ is the nearest-neighbor hopping amplitude between $t_{2g}$ orbitals,
$E_1=U-3J_H$, $E_2=U-J_H$, $E_3=U+2J_H$, where $U$ and $J_H$ are Coulomb and
Hund's interactions, respectively. The orbital operators
$\mathcal{O}^{(\gamma)}_{ij}$, $\mathcal{P}^{(\gamma)}_{ij}$, and $n_i^{(\gamma)}$
depend on $\vec {r}_{ij}$-bond direction $\gamma$. In systems with strong SOC,
it is convenient to represent these operators in terms of orbital angular
momentum operators $l_{x,y,z}$ of $t_{2g}$ level. For $\gamma=x$, we have
\begin{gather}
\mathcal{O}^{(x)}_{ij}= [(1-l^2_y)_i(1-l^2_y)_j+(l_y l_z)_i(l_z l_y)_j]
+[y\leftrightarrow z], \notag \\
\mathcal{P}^{(x)}_{ij}= [(1-l^2_y)_i(1-l^2_y)_j+(l_y l_z)_i(l_y l_z)_j]
+[y\leftrightarrow z], \notag \\
n^{(x)}=l^2_x,
\end{gather}
while the corresponding expressions for $\gamma=y$ bond follow from
symmetry.

Next, we project the Hamiltonian of Eq.~\ref{eq:H1} onto pseudospin-1/2
subspace defined by wavefunctions $|\widetilde{A}_{\pm}\rangle$ (Eq.~2 in the
main text). In terms of orbital and spin quantum numbers $|l_z,s_z\rangle$,
they read as follows:
\begin{align}
\label{eq:gs}
|\widetilde{A}_+\rangle &= \frac{1}{\sqrt{1\!+\!|\eta_+|^2}}
\left(\sin\theta\;|0,+\tfrac{1}{2}\rangle-
\cos\theta\;|+1,-\tfrac{1}{2}\rangle + \eta_+|-1,-\tfrac{1}{2}\rangle\right),
\notag \\
|\widetilde{A}_-\rangle &= \frac{1}{\sqrt{1\!+\!|\eta_{-}|^2}}
\left(\sin\theta\;|0,-\tfrac{1}{2}\rangle-
\cos\theta\;|-1,+\tfrac{1}{2}\rangle + \eta_{-}|+1,+\tfrac{1}{2}\rangle\right).
\end{align}
We recall $\eta_{\pm}=\frac{\cos\theta}{E_{BA}}
(\pm ig_1\varepsilon_1+g_2\varepsilon_2)$. The pseudospin-lattice coupling
$\mathcal{H}_{\rm s-l}$ that we are looking for will be generated by the
$\eta_{\pm}$-corrections to $|\widetilde{A}_{\pm}\rangle$.

To project Eq.~\ref{eq:H1} onto pseudospin $S=1/2$ sector, one has to
calculate the matrix elements of various combinations of spin $s_\alpha$ and
orbital $l_\alpha$ operators between $|\widetilde{A}_{\pm}\rangle$ states, and
express them via pseudospin $S=1/2$. For example,
\begin{align}
&\langle\widetilde{A}_+|s^+|\widetilde{A}_+\rangle  =0 ,
\ \ \ \ \ \ \ \ \ \ \ \ \ \ \ \ \ \ \ \
\langle\widetilde{A}_+|s^+|\widetilde{A}_-\rangle
 =\sin ^2 \theta,
\notag \\
& \langle\widetilde{A}_-|s^+|\widetilde{A}_+\rangle
=-2 \eta_+ \cos \theta ,
\; \ \ \ \ \ \ \
\langle\widetilde{A}_-|s^+|\widetilde{A}_-\rangle =0 ,
\end{align}
to first order in $\eta_{\pm}$. Thus,
$s^+=\sin ^2 \theta \;S^+-2\eta_+\cos\theta \;S^- $.
Along the same lines, one obtains the other operators that enter
in Eq.~\ref{eq:H1} and contain $\eta_{\pm}$-terms of interest:
\begin{align}
s^+ l_x^2 &=\sin^2 \theta\; S^+ + \tfrac{1}{2} \cos^2 \theta\; S^-
-\eta_+ \cos\theta\; S^- \; , \notag \\
s^+ l_z^2 &=-2\eta_+ \cos\theta\; S^- \; , \notag \\
l_z l_x &=\tfrac{\sin \theta}{\sqrt{2}}
\left[
(\cos\theta + \eta_+)\; S^- - (\cos\theta + \eta_-)\; S^+
\right] , \notag \\
s^z l_z l_x &=\tfrac{\sin \theta}{2\sqrt{2}}
\left[
(\cos\theta + \eta_+)\; S^- + (\cos\theta + \eta_-)\; S^+
\right] .
\end{align}
Collecting all the terms $\propto\eta_{\pm}$, we find the pseudospin-lattice
coupling [Eq.~(4) of the main text]:
\begin{equation}
\label{eq:s-lsM}
\mathcal{H}_{\rm{s-l}}^{ij}\!=\!
\tilde{g}_1 \varepsilon_1\;(S_i^xS_j^y\!+\!S_i^yS_j^x)+
\tilde{g}_2 \varepsilon_2\;(S_i^xS_j^x\!-\!S_i^yS_j^y),
\end{equation}
where $\tilde{g}\!=\!\kappa g$ and
$\kappa\!\simeq\!\frac{t^2}{U}\frac{\sin^2 2\theta}{E_{BA}}\frac{J_H}{U}$
(to first order in Hund's coupling). In principle, the form of this
Hamiltonian can be inferred from the symmetry considerations (see discussion
in the main text).

\section{B. Mean-field ground state of $J_{\rm{eff}}=0$ model
under tetragonal field}

A standard way of handling a hard-core constraint $n^T_i\leq 1$ in
singlet-triplet models (see, e.g., Refs.~\cite{Som01,Mat04}) is to represent
$T_\alpha$-boson via the singlet $s$ and triplet $t_\alpha$ particles:
$T_{i,\alpha}^\dagger \Rightarrow t_{i,\alpha}^\dagger s_i$.
In this representation,
$T_{i,\alpha}^\dagger T_{i,\alpha}=t_{i,\alpha}^\dagger t_{i,\alpha}$, while
$T_{i,\alpha}^\dagger T_{j,\alpha}=s_j^{\dagger} s_it_{i,\alpha}^\dagger t_{j,\alpha}$.
A particle number constraint
$s_i^{\dagger} s_i+\sum_{\alpha} t_{i,\alpha}^\dagger t_{i,\alpha}=1 $
is implied. On a mean-field level, a singlet amplitude $s$ is replaced by its
classical average, $s_i\approx \sqrt{1-n}$~, where
$n=\sum_{\alpha} \langle t_{i,\alpha}^\dagger t_{i,\alpha}\rangle$,
resulting in Gutzwiller-type renormalization of the intersite terms
$T_{i,\alpha}^\dagger T_{j,\alpha}\approx (1-n)\;t_{i,\alpha}^\dagger t_{j,\alpha}$
in the exchange $J$-Hamiltonian.

Evaluation of a mean-field ground state energy is then straightforward. One
first obtains a single-ion multiplets (see Fig.~S1), quantified by the lowest
singlet $s$-level at energy
\begin{equation}
E_s(\delta)=\left(\frac{3}{2}-\frac{\delta}{3}-\sqrt{\frac{9}{4}
-\delta+\delta^2} \right)\lambda \;,
\end{equation}
a distance to the $xy$-doublet
\begin{equation}
E(\delta)=\left(\frac{1}{2}+\sqrt{\frac{9}{4}-\delta+\delta^2}-
\sqrt{1+\delta^2}\right)\lambda\;,
\end{equation}
and splitting of a triplet level by
$\Delta_z(\delta)=(\delta+\sqrt{1+\delta^2}-1)\lambda$, where
$\delta=g\varepsilon/2\lambda$. On a classical level, the high energy $T_z$
state is irrelevant, and the exchange interactions induce a condensation of
$T_{x/y}$~-type bosons, forming a magnetic quadrupolar moment
$Q_T=n^T_x+n^T_y-2n^T_z$ in the ground state. The value of $Q_T$ and
tetragonal distortion $\varepsilon$ are obtained by minimizing the
total energy including spin-orbit and Jahn-Teller couplings, exchange
interactions, and elastic energy:
\begin{equation}
E_{\rm {total}}=-2JQ_T^2(\delta) +E_s(\delta)+\tfrac{1}{2}K\varepsilon^2 \;,
\end{equation}
where
\begin{equation}
Q_T(\delta)=\tfrac{1}{2}[1-\tfrac{E(\delta)}{2J}].
\end{equation}
A numerical analysis of the above equations results in a phase diagram
shown in Fig.~4(b) of the main text. The main effect of JT coupling is to
reduce spin-orbit coupling induced gap $\lambda$ to a smaller value
of $E(\delta)$, promoting thereby $XY$-type magnetic condensate.

\begin{figure}
\begin{center}
\includegraphics[width=9cm]{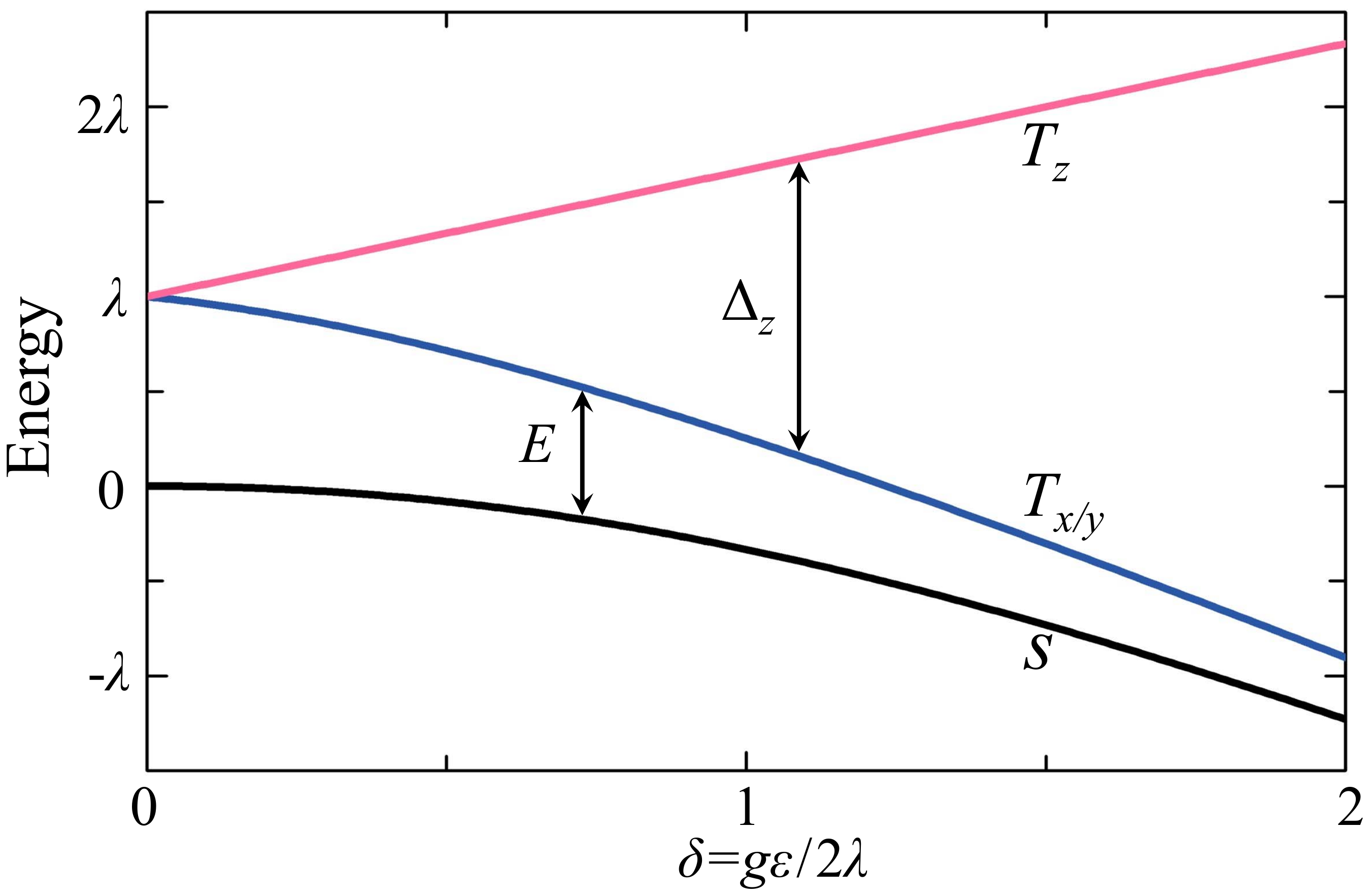}
\caption{The energy levels of Ru$^{4+}$ ion under a combined action of
spin-orbit coupling $\lambda$ and Jahn-Teller field $g\varepsilon$ of a
tetragonal symmetry. Distance $E$ from a ground state singlet $s$ (black) to
$T_{x/y}$ doublet (blue) defines the spin gap to be overcomed by the exchange
field $J$. Above a critical value $J_{cr}\sim E/2$, the $T_{x/y}$ bosons condense,
while upper level $T_z$ (red) remains unpopulated.}
\label{fig:s2}
\end{center}
\end{figure}


\begin{thebibliography}{99}
\bibitem{notePJT}
The pseudo-JT effect is an extension of the JT effect to systems with
orbital-singlet ground states; it operates through the mixing of the ground
and excited states under the nuclear displacements~\cite{Ber13,Ber06}.
\bibitem{noteAB} In a solid, spin-orbit excitation energy $E_{BA}$ depends
on momentum~\cite{Kim14,Plo16}; we neglect this effect here.
\bibitem{noteK} On a classical level, i.e., neglecting quantum corrections
from magnon-magnon interactions, in-plane anisotropy is solely due to
spin-lattice coupling.
\bibitem{noteCo}
We recall that $\varepsilon\propto \tilde g \propto 1/\lambda$.
In pseudospin-1/2 Co$^{2+}$ compounds, where $\lambda$ is one order of value
smaller than in iridates, magnetic order induced distortions should be
of the order of $10^{-3}$, as indeed observed~\cite{Oka61,Tom50}.
\end{thebibliography}
\end{document}